\documentstyle[emulateapj]{article}

\newcommand\Ha{H$\alpha$}

\newcommand\hii{\ion{H}{2}}
\newcommand\hi{\ion{H}{1}}

\newcommand\mbol{$M_{\rm bol}$}
\newcommand\teff{$T_{\rm eff}$}

\received{10-August-2004}
\accepted{21-September-2004}

\slugcomment{Accepted to the {\bf ASTRONOMICAL JOURNAL} 21 September 2004}

\lefthead{Oey, Watson, Kern, and Walth}
\righthead{HIERARCHICAL TRIGGERED STAR FORMATION}

\begin{document}

\title{
Hierarchical Triggering of Star Formation by Superbubbles in W3/W4}

\author{M. S. Oey}
\affil{Department of Astronomy, 830 Dennison Building, 
	University of Michigan, Ann Arbor, MI\ \ \ 48109-1090}
\author{Alan M. Watson}
\affil{Centro de Radiostronom\'\i a y Astrof\'\i sica, Universidad
        Nacional Aut\'onoma de M\'exico, Apartado Postal 3-72, 58089
	Morelia, Michoac\'an, M\'exico} 
\smallskip
\author{Katie Kern\altaffilmark{1}}
\affil{Dept. of Astronomy, University of Wisconsin, 475 Charter St., 
	Madison, WI\ \ \ 53706}
\author{Gregory L. Walth}
\affil{Lowell Observatory, 1400 W. Mars Hill Rd., 
	Flagstaff, AZ\ \ \ 86001}

\altaffiltext{1}{Participant in the 2003 Research Experience for
	Undergraduates Program, Northern Arizona University.}

\begin{abstract}

It is generally believed that expanding superbubbles and mechanical
feedback from massive stars trigger star formation, because there are
numerous examples of superbubbles showing secondary star formation at 
their edges.  However, while these systems show an age sequence, they
do not provide strong evidence of a causal relationship.  The W3/W4
Galactic star-forming complex suggests a three-generation hierarchy:
the supergiant shell structures correspond to the oldest
generation; these triggered the formation of IC~1795 in W3, the
progenitor of a molecular superbubble; which in turn triggered the
current star-forming episodes in the embedded regions W3-North,
W3-Main, and W3-OH.  We present $UBV$ photometry and spectroscopic
classifications for IC~1795, which show an age of 3 -- 5 Myr.  This
age is intermediate between the reported 6 -- 20 Myr age of the
supergiant shell system, and the extremely young ages
($10^4 - 10^5$ yr) for the embedded knots of ultracompact \hii\
regions, W3-North, W3-Main, and W3-OH. 
Thus, an age sequence is indeed confirmed for the entire W3/W4
hierarchical system.  This therefore provides some of the first
convincing evidence that superbubble action and mechanical feedback are
indeed a triggering mechanism for star formation.

\end{abstract}


\section{Introduction}

The shocks in expanding superbubbles and supernova remnants are widely
believed to be a primary mechanism for triggering star formation.
Although this mechanical feedback from massive stars cannot be the
only catalyst for the gravitational collapse of molecular clouds,
there are numerous examples of superbubbles that show young,
star-forming regions at their edges.  In the Large Magellanic Cloud,
superbubble systems like N11 (Walborn \& Parker 1992), N44 (Oey \&
Massey 1995), and N51 D (Oey \& Smedley 1998) all show young OB
associations, typically with ages $\lesssim 3$ Myr, on the outer edges
of the shells, while associations that are somewhat older, typically
by a few Myr, are responsible for the formation of the shells themselves.  

However, it is important to distinguish between
{\it sequential} star formation and {\it triggered} star formation.
Studies of the stellar populations in systems like the above examples
conclusively demonstrate that the star formation is sequential,
but it is much more difficult to demonstrate a causal relationship.
For any star formation that takes place in two adjacent molecular
clouds, or even subregions of the same cloud, if the
superbubble from the older region grows to impact the younger region,
the resulting configuration will show a younger nodule on the edge of
an older one.  Thus, although they are suggestive of triggering,
two-generation systems do not constitute strong evidence that it is
actually taking place.

On the other hand, a hierarchical system of three or more generations
would provide much stronger evidence of a causal relationship.  An
example of superbubble-triggered star formation over three generations
should show the oldest shell with a younger superbubble at its edge,
which in turn shows even younger star formation on its edge.  Such a
scenario is extremely unlikely to be coincidental, sequential star
formation.  In what follows, we identify the W3/W4 Galactic star-forming
complex as the first example of three-generation, hierarchical triggered
star formation.  Thus this offers much more concrete support for the
widely-held view that superbubble expansion triggers star formation.

\begin{figure*}
\epsscale{1.0}
\plotone{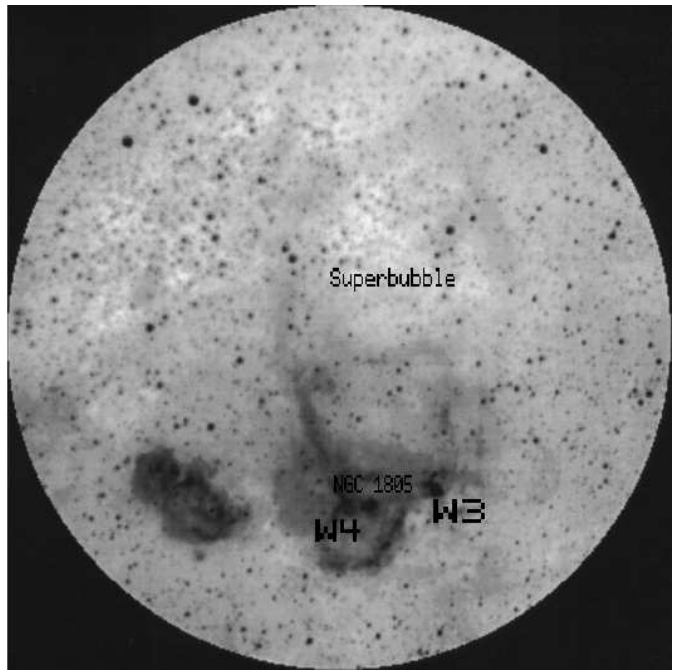}
\caption{\Ha\ image of the $\sim$1300-pc superbubble that is blowing out
above the Galactic plane from the W4 region (image from Dennison et
al. 1997).  
\label{f_dennison}}
\end{figure*}

\begin{figure*}
\epsscale{1.0}
\caption{Grayscale of the W3 region from the DSS,
overlaid by CO contours from the FCRAO CO survey.
[{\tt Figure available as JPEG on the astro-ph submission.}]
\label{f_w3}}
\end{figure*}

\section{The W3/W4 Complex}

The W3/W4 complex is located in the Perseus arm of the Galaxy, at a
distance of about 2.3 kpc (Massey et al. 1995a).  W4 shows a nebular
shell roughly $1^\circ$ in diameter, and it is also at the base of the
well-known Perseus ``chimney'' that is seen in \hi\ (Normandeau et
al. 1996).  Dennison et al. (1997) find that the ``chimney'' is the
lower section of an elongated, but closed, shell that extends
$\sim$230 pc above the Galactic plane and is visible in \Ha\
(Figure~\ref{f_dennison}).  Most recently, Reynolds et al. (2001)
identify an even larger loop that is $\sim$1300 kpc above 
the Galactic plane, of which the Dennison et al. superbubble is 
the lower section.  For all of these shell features, the W4 region
is located at the Galactic plane apex.  At the center of the W4 shell
is the OB association IC~1805.  

W3 (Figure~\ref{f_w3}) is a smaller \hii\ region, $\sim 30\arcmin$
(20 pc) in diameter, on the northwest edge of the W4 shell.  As a
thermal radio source, W3 subdivides into a complex with three dominant
knots of ultracompact \hii\ regions:  W3-Main, W3-North, and W3-OH.
As seen in Figure~\ref{f_w3}, these knots are embedded in 
a shell of molecular gas that surrounds the OB association IC~1795.
The young, continuum radio-emitting cluster NGC 896 is seen against
the western edge of the CO shell.

The configuration of the larger Perseus chimney/superbubble and
smaller IC~1795/W3 superbubble are strongly suggestive of a
three-generation system of hierarchical triggered star formation.
That scenario would imply that the Perseus chimney/superbubble, 
formed by the first generation, triggered the formation of
IC~1795 as the second generation, and its superbubble in turn
triggered the W3 embedded star-forming regions W3-Main, W3-North, and
W3-OH, thereby constituting the third generation.  If this model
is correct, it should be reflected by a corresponding age sequence.
In this study, we evaluate the ages of the associated stellar
populations for these three putative generations, based on both new
data for IC~1795 and studies in the literature.  

\section{IC~1795}

Since IC~1795 represents the intermediate, second generation of stars
in the complex, strong constraints on its age are necessary to test
the model of hierarchical triggering of star formation.  To date, the
only published studies of the stellar population in 
IC~1795 are based on photographic and photoelectric studies by
Ogura \& Ishida (1976) and Sato (1970).  We have
therefore carried out a new investigation based on modern CCD
photometry and spectroscopic observations.  While our intent was to
examine IC~1795 in detail, we also obtained stellar data for the
wider W3 region shown in Figure~\ref{f_mosaic}; these data are
presented in the Appendix.

\subsection{Photometric observations}


\begin{figure*}
\epsscale{2.0}
\plotone{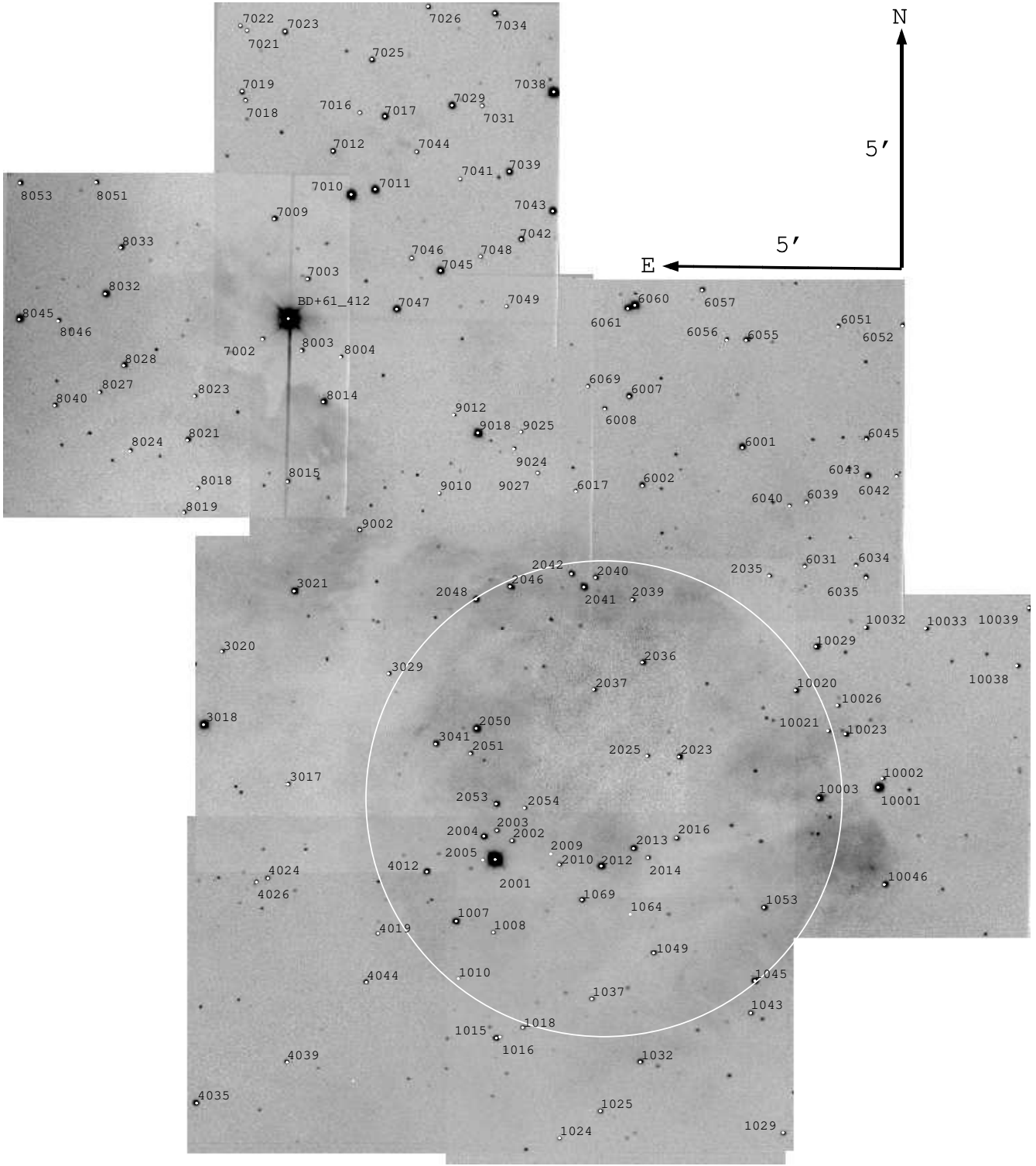}
\caption{ $U$-band mosaic of the W3 region, with stars having new $U$-band
photometry identified, for $U<18$.  The circle shows the IC~1795 region for which
the H-R diagram was constructed. 
\label{f_mosaic}}
\end{figure*}

{\itshape UBV} imaging was obtained on 1999 October 6--9 at the 84-cm
telescope of the Observatorio Astron\'omico Nacional at San Pedro
M\'artir (SPM), Baja California Norte, M\'exico. We used a $1024 \times
1024$ SITe CCD binned $2 \times 2$, which gave $0.85\arcsec$ pixels and an
instantaneous field of view of $7\arcmin \times 7\arcmin$.  The seeing was 
typically around $2\arcsec$ and the sky was clear.  IC 1795 was imaged in four
pointings, with BD~+61~411 in each corner of the CCD.  NGC 896 and
W3-Main were imaged in one pointing each, and W3-North was imaged in
three pointings.  Our $U$-band mosaic of the entire region is shown in
Figure~\ref{f_mosaic}.  At each pointing, we obtained
exposures of 600 seconds in {\itshape V}, 900 seconds in {\itshape B},
and 1200 seconds in {\itshape U}, along with shorter 10 and 100 second
images in each filter.  Each image was corrected for bias offset and was
then divided by a twilight flat field.  Variance between the flat fields
obtained each nights and direct measurements of a grid stars suggest
that the flat fields are good to about 0.5\% rms.

We observed roughly 15 Landolt (1983) standards each night over a range
of airmass and color. Instrumental magnitudes were determined by
synthetic aperture photometry with a $17\arcsec$ star aperture and a sky
annulus with inner and outer diameters of $34\arcsec$ and $51\arcsec$. We
determined the transformation between the instrumental and standard
systems by fitting all of the $\sim$60 observations simultaneously, with a
separate linear extinction coefficient for each night, but a single set
of color coefficients and zero points. Linear color transformations were
adequate for {\itshape B} and {\itshape V}, but the data required a
quadratic color transformation for {\itshape U}. The RMS residuals were
2.1\% in {\itshape U} and 1.5\% in {\itshape B} and {\itshape V}.

We selected stars by eye in each 600 second {\itshape V}\ image, as
these were by far the deepest images. We then performed synthetic
aperture photometry on each image using an $8.5\arcsec$ diameter star
aperture and a sky annulus with inner and outer diameters of
$34\arcsec$ and $51\arcsec$.  Aperture corrections between $8.5\arcsec$
and $17\arcsec$ diameter apertures were 
determined for each image.  We then calculated the standard magnitude for
each star and its uncertainty, taking into account read noise and photon
noise in the star and sky apertures, flat-field uncertainties, and
transformation uncertainties.  Since we had three (10 second, 100
second, and 600/900/1200 second) images in each band, we selected the
unsaturated measurement with the lowest uncertainty.  Our complete set
of new $UBV$ photometry is presented in the Appendix,
Table~\ref{t_photall}.

\begin{figure*}
\epsscale{1.0}
\plotone{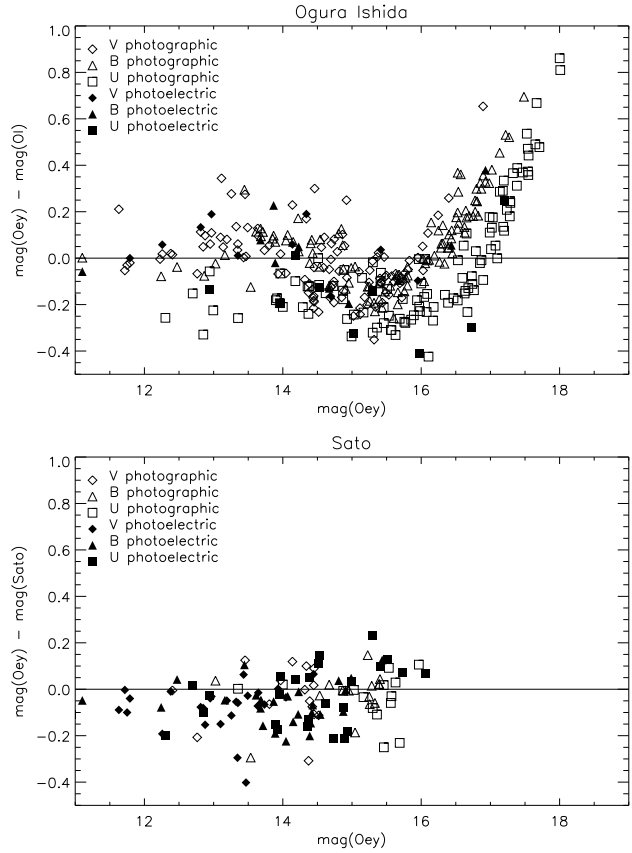}
\caption{ Photometric differences of our photometry with those of
Ogura \& Ishida (1976; top) and Sato (1970; bottom), as a function of
magnitude. 
\label{f_photcomp}}
\end{figure*}

We compared our data with photometry reported by Ogura \& Ishida
(1976) and Sato (1970).  The Ogura \& Ishida measurements are mostly
based on iris photometry of photographic plates, calibrated to a
subset of photoelectric data, while those of Sato are photoelectric.
Figure~\ref{f_photcomp} shows the comparison of our modern data to
these earlier studies.  There is a large scatter in the photometric
differences for both, with a large systematic pattern seen in the
comparison to the data of Ogura \& Ishida as function of magnitude.
The differences cannot be explained with the published information.
One might suspect a non-linearity in the detector, but thorough annual
``health checks'' of this CCD performed every year since 2000 have
failed to find any evidence for such a problem.  We note that
Figure~\ref{f_photcomp} shows that the photometry from these previous
studies also appear to be inconsistent with each other. 

\subsection{Spectroscopic observations}

Spectroscopic observations of candidates for the earliest-type stars
were obtained with the 2.1-m telescope at SPM during the nights of 1999 
October 15 -- 17.  We used the 600 line mm$^{-1}$ grating on the Boller
\& Chivens spectrograph and Thomson 2048 $\times$ 2048 CCD, yielding a
useable wavelength coverage of about 3500 -- 5900 \AA\ at spatial
and spectral resolutions of about 3\AA\ and 1.3$\arcsec\ \rm px^{-1}$,
respectively.  The stars
observed spectroscopically were selected from the wider region around
IC~1795, not just in the central region that we imaged in $UBV$.
The spectroscopic candidates were the bluest by their reddening-free
$Q$ values ($Q \lesssim -0.3$) and brightest ($V \lesssim 15)$ as
reported by Ogura \& Ishida (1976) and Sato (1970).

The spectroscopic data were reduced with the {\sc iraf}\footnote{{\sc
iraf} is distributed by NOAO, which is operated by AURA, Inc., under
cooperative agreement with the NSF.} software.  The raw data were
bias-subtracted and then flat-fielded using the quartz lamp
calibration images.  Owing to a problem with the sky frames, an
illumination correction was only applied to data taken on 1999 October
15.  During aperture extraction, the local background was evaluated for
every set of 10 rows in the chip and subtracted; since this was
straightforward, the lack of an explicit illumination correction is
considered unimportant.  The spectra were extracted following standard
procedures by fitting the trace along the dispersion and applying
wavelength calibrations from He-Ne-Ar arc lamp observations taken
during each night.

The twenty stars with spectroscopic observations were independently
classified by KK and MSO, with agreement generally within one spectral
subtype, and differences resolved upon reinspection.  O and early
B-type stars were classified according to the criteria of Walborn \&
Fitzpatrick (1990).  Our observations of spectroscopic standards aided
in classifying the stars. 

\subsection{H-R Diagram}

We focused on the stellar population of IC~1795, taken to be the stars
within a $5\arcmin$-radius, circular region centered at $02^h 26^m
15.^s0, +62^\circ 02\arcmin 00.\arcsec 0$ (J2000.0) (Figure~\ref{f_mosaic}). 
The stellar effective temperatures \teff\ and bolometric corrections (BCs)
were estimated for the stars with spectral classifications using the
calibrations of Chlebowski \& Garmany (1991) for O stars and Humphreys
\& McElroy (1984) for B stars.  For the dominant, bright star BD +61
411 (ID number 2001 = OI 89) we adopt spectral type O6.5 V((f)) from
Mathys (1989).  For blue stars (reddening-corrected $B-V < -0.22$ or
$Q < -0.6$) without direct spectroscopic 
classifications, but having good photometry with formal errors in
$V,\ B-V,$ and $U-B$ less than 0.03, 0.05, and 0.05, respectively, we
estimated log \teff\ from the reddening-free parameter $Q = (U-B) - 0.72
(B-V)$ using the relations of Massey et al. (1995b).  These authors
also derive a relation between the BC and log \teff, which we then used to
obtain the BCs for these stars.  The remaining stars have poorer
photometry, and for these we estimated log \teff\ using only the
reddening-corrected $(B-V)_0$, from the relation also given by Massey
et al. (1995b).  We used a fixed value for the color excess of $E(B-V)
= 1.25$, which is the median value for the stars with spectroscopic
classifications.  The same relation between BC and log \teff\
was then applied to obtain the bolometric corrections.

With the bolometric corrections in hand, we then computed the
bolometric magnitudes, as usual correcting for extinction and distance,
as \mbol$ = V - A_V + \rm BC +DM$.  Hanson \& Clayton (1993)
determine $A_V = 2.9\ E(B-V)$ for the neighboring region IC 1805, so
we adopt this relation between the total to selective extinction.
Likewise, we adopt the distance modulus $\rm DM=11.85$ (2.3 kpc)
derived by Massey et al. (1995a) for IC 1805, a value consistent with
the distance adopted in past studies of the W3 complex.

\begin{figure*}
\epsscale{1.5}
\plotone{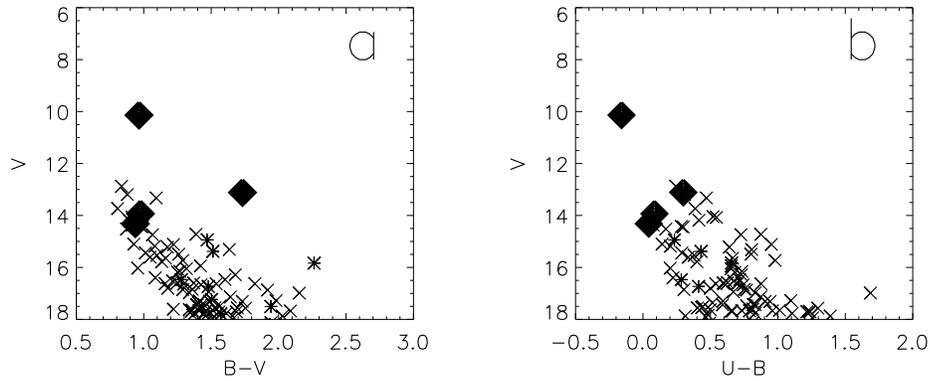}
\vspace*{-4.5 truein}
\caption{Color-magnitude diagrams for IC~1795.  Solid symbols and
asterisks show, respectively, stars with spectroscopic
classifications and reliable reddening-free $Q$ values; crosses show
the remainder, lower-quality data (see text).  Star 3041 (OI 109; the 
second-brightest star) is heavily reddened and appears 
displaced relative to the locus of the other stars. 
\label{f_cmds}}
\end{figure*}

\begin{figure*}
\epsscale{2.0}
\plotone{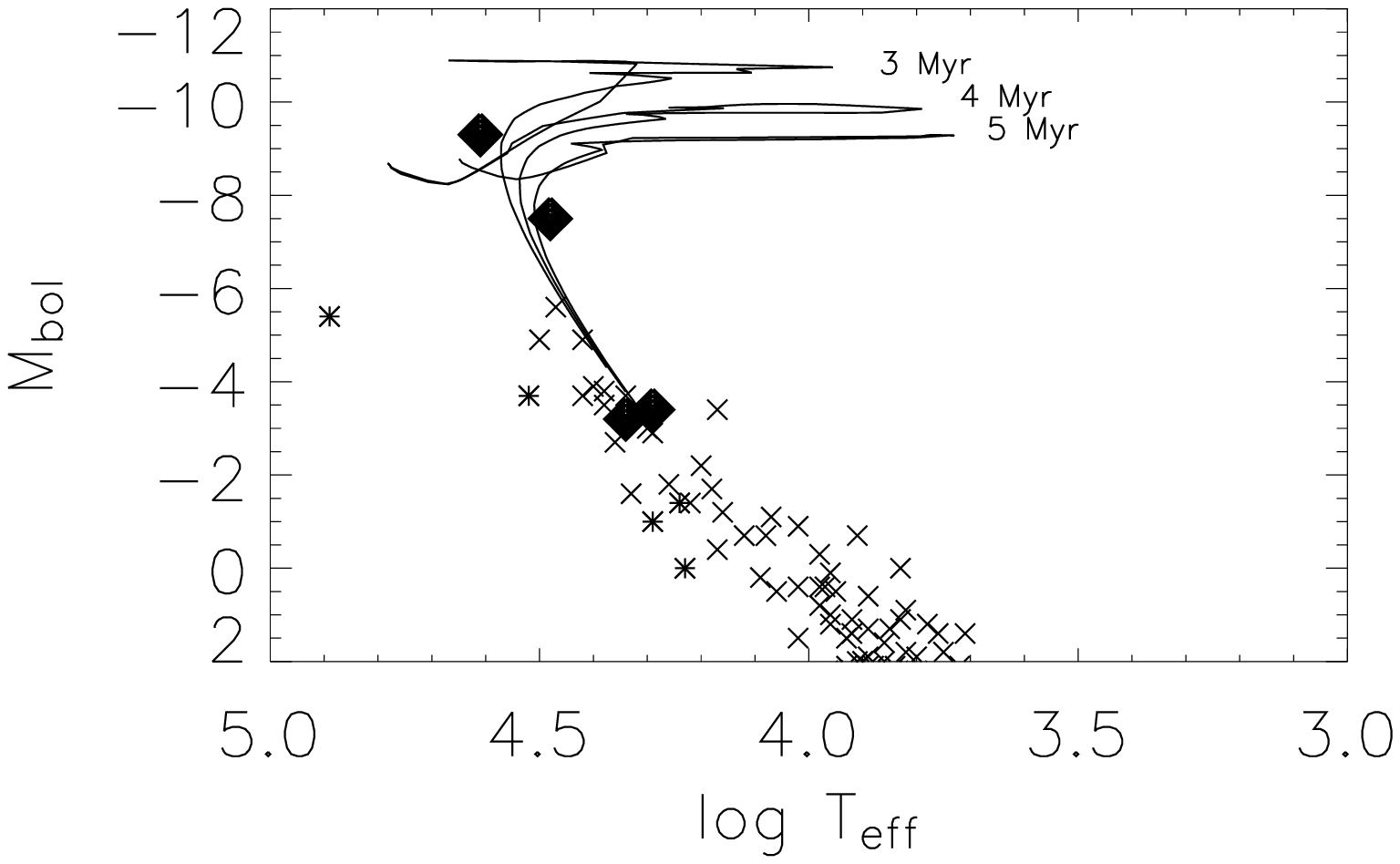}
\vspace*{-4 truein}
\caption{H-R diagram for IC~1795.  Solid, asterisk, and cross symbols show
stars with \teff\ and \mbol\ estimated from, respectively, spectroscopic
classifications, reddening-free $Q$ values, and $(B-V)_0$ only.
Isochrones for 3, 4, and 5 Myr from Schaller et al. (1992) are
overplotted as labeled.
\label{f_hrd}}
\end{figure*}

Figures~\ref{f_cmds}$a$ and $b$ show the $B-V$ vs $V$ and $U-B$ vs $V$
color-magnitude diagrams for IC~1795, respectively, while
Figure~\ref{f_hrd} shows the resulting \mbol\ vs \teff\ H-R diagram.
Stars 1053 (OI 42), 2004 (OI 86), and 10003 (OI 40) were omitted since
their spectral types revealed them to be foreground F and G dwarfs.
Solid symbols represent stars placed on the basis of spectroscopic
classifications; asterisks represent the stars with good photometry
that are placed by their $Q$ values; and crosses represent the
remainder of the data, placed by their $(B-V)_0$ colors alone.
Isochrones for 3, 4, and 5 Myr are overplotted in Figure~\ref{f_hrd}, from
Schaller et al. (1992) for solar metallicity.  From Figure~\ref{f_hrd}, we
infer an age of 3 -- 5 Myr for IC~1795.  Data for the individual stars
in IC~1795 are presented in Table~\ref{t_phot}:  The first two columns
give ID numbers assigned by this work and Ogura  
\& Ishida (1976), respectively, columns 3 and 4 give the celestial
coordinates for epoch J2000.0, and columns 5 -- 10 list the $V$, $B$, and
$U$ magnitudes with corresponding formal (non-systematic)
uncertainties.  Column 11 gives our spectral classification for
stars that were observed spectroscopically.

\begin{deluxetable}{rrccccccccl}
\scriptsize
\tablewidth{0pt}
\tablecolumns{11}
\tablecaption{Photometry and Spectral Types for IC 1795 \label{t_phot}}
\tablehead{
\colhead{ID} & {OI} & {RA(J2000.0)} & {Dec(J2000.0)} &
\colhead{$V$} & {$V$ err} & \colhead{$B$} & {$B$ err} & 
\colhead{$U$} & {$U$ err} & Sp Type }
\startdata
1007 & 112 & 02:26:41.31 & +61:59:24.2 & 14.318 & 0.001 & 15.256 & 0.001 & 15.299 & 0.002 & B1-2 V \\
1008 &\nodata & 02:26:34.63 & +61:59:10.6 & 15.312 & 0.001 & 16.947 & 0.003 & 17.753 & 0.011 &\nodata \\
1009 &\nodata & 02:26:38.45 & +61:59:26.2 & 18.138 & 0.013 & 19.585 & 0.028 & 20.080 & 0.120 &\nodata \\
1010\tablenotemark{a} &\nodata & 02:26:40.83 & +61:58:11.7 & 17.233 & 0.005 & 18.736 & 0.011 & 18.000 & 0.023 &\nodata \\
1011 &\nodata & 02:26:34.54 & +61:58:16.6 & 18.124 & 0.013 & 20.069 & 0.042 & 20.883 & 0.235 &\nodata \\
1034 &\nodata & 02:26:26.63 & +61:57:41.9 & 17.292 & 0.006 & 18.705 & 0.013 & 19.298 & 0.055 &\nodata \\
1035 &\nodata & 02:26:27.21 & +61:58:14.4 & 17.278 & 0.006 & 19.260 & 0.018 & 20.356 & 0.147 &\nodata \\
1036 &\nodata & 02:26:21.67 & +61:57:55.0 & 16.840 & 0.004 & 18.199 & 0.008 & 18.951 & 0.039 &\nodata \\
1037 &\nodata & 02:26:16.89 & +61:57:47.9 & 15.778 & 0.002 & 16.913 & 0.004 & 17.313 & 0.010 &\nodata \\
1038 &\nodata & 02:26:17.89 & +61:57:25.3 & 17.691 & 0.008 & 19.148 & 0.018 & 20.382 & 0.150 &\nodata \\
1039 &\nodata & 02:26:09.30 & +61:57:32.1 & 17.434 & 0.006 & 18.826 & 0.013 & 19.659 & 0.073 &\nodata \\
1040 &\nodata & 02:26:10.50 & +61:58:02.0 & 16.733 & 0.004 & 18.212 & 0.008 & 18.624 & 0.030 &\nodata \\
1045 &\nodata & 02:25:47.67 & +61:58:12.6 & 14.184 & 0.001 & 15.076 & 0.001 & 15.491 & 0.003 &\nodata \\
1046 &\nodata & 02:25:56.53 & +61:58:01.9 & 17.430 & 0.006 & 18.875 & 0.014 & 19.693 & 0.083 &\nodata \\
1047 &\nodata & 02:26:00.88 & +61:58:15.9 & 17.550 & 0.008 & 19.079 & 0.019 & 19.514 & 0.075 &\nodata \\
1048 &\nodata & 02:26:04.52 & +61:58:26.8 & 16.986 & 0.006 & 19.140 & 0.020 & 20.827 & 0.254 &\nodata \\
1049 &\nodata & 02:26:05.86 & +61:58:46.6 & 14.939 & 0.001 & 16.410 & 0.002 & 16.643 & 0.006 &\nodata \\
1050 &\nodata & 02:26:07.08 & +61:58:41.7 & 17.712 & 0.008 & 19.269 & 0.019 & 20.507 & 0.163 &\nodata \\
1051 &\nodata & 02:25:58.48 & +61:58:49.5 & 17.481 & 0.008 & 19.047 & 0.019 & 19.816 & 0.095 &\nodata \\
1052 &\nodata & 02:25:46.07 & +61:59:10.6 & 17.624 & 0.008 & 18.963 & 0.017 & 19.462 & 0.072 &\nodata \\
1053\tablenotemark{b} & 42 & 02:25:46.12 & +61:59:45.2 & 14.447 & 0.001 & 15.602 & 0.001 & 16.180 & 0.004 & G3 V   \\
1054 &\nodata & 02:25:51.04 & +61:59:54.3 & 16.872 & 0.004 & 18.790 & 0.012 & 19.558 & 0.062 &\nodata \\
1059 &\nodata & 02:25:59.35 & +62:01:09.4 & 18.190 & 0.021 & 20.167 & 0.058 & 22.365 & 1.224 &\nodata \\
1061 &\nodata & 02:26:03.27 & +61:59:36.1 & 16.641 & 0.004 & 17.937 & 0.007 & 18.635 & 0.031 &\nodata \\
1062 &\nodata & 02:26:04.96 & +61:59:40.1 & 17.873 & 0.010 & 19.902 & 0.034 & 21.290 & 0.348 &\nodata \\
1063 &\nodata & 02:26:06.20 & +61:59:19.2 & 17.871 & 0.010 & 19.329 & 0.022 & 19.643 & 0.076 &\nodata \\
1064\tablenotemark{a} &\nodata & 02:26:10.09 & +61:59:35.0 & 17.791 & 0.009 & 19.363 & 0.022 & 16.960 & 0.015 &\nodata \\
1065 &\nodata & 02:26:05.55 & +62:00:30.0 & 17.940 & 0.014 & 19.610 & 0.031 & 20.867 & 0.226 &\nodata \\
1069 & 92 & 02:26:18.80 & +61:59:52.9 & 15.182 & 0.001 & 16.264 & 0.002 & 16.467 & 0.005 &\nodata \\
1070 &\nodata & 02:26:25.07 & +61:59:38.4 & 17.362 & 0.007 & 18.768 & 0.014 & 19.283 & 0.055 &\nodata \\
2001 & 89 & 02:26:34.41 & +62:00:42.6 & 10.137 & 0.000 & 11.102 & 0.001 & 10.943 & 0.000 & O6.5 V((f)) \tablenotemark{c}  \\
2002 & 87 & 02:26:31.43 & +62:01:06.7 & 15.532 & 0.002 & 16.630 & 0.003 & 17.009 & 0.007 &\nodata \\
2003 & 85 & 02:26:34.15 & +62:01:19.3 & 14.725 & 0.001 & 16.113 & 0.001 & 16.987 & 0.005 &\nodata \\
2004\tablenotemark{b} & 86 & 02:26:36.47 & +62:01:11.9 & 13.968 & 0.001 & 15.044 & 0.001 & 15.458 & 0.002 & F8 V   \\
2005\tablenotemark{a} &\nodata & 02:26:36.64 & +62:00:41.8 & 16.285 & 0.005 & 16.098 & 0.002 & 13.596 & 0.002 &\nodata \\
2006 &\nodata & 02:26:38.83 & +62:00:30.4 & 16.470 & 0.003 & 18.073 & 0.007 & 18.791 & 0.028 &\nodata \\
2007 &\nodata & 02:26:29.62 & +62:00:19.1 & 17.768 & 0.011 & 19.380 & 0.021 & 20.486 & 0.146 &\nodata \\
2008 &\nodata & 02:26:26.17 & +62:00:32.5 & 17.708 & 0.012 & 19.395 & 0.028 & 20.611 & 0.174 &\nodata \\
2009\tablenotemark{a} &\nodata & 02:26:24.43 & +62:00:50.3 & 17.504 & 0.008 & 19.447 & 0.024 & 16.589 & 0.015 &\nodata \\
2010 & 90 & 02:26:22.89 & +62:00:37.1 & 15.383 & 0.002 & 16.896 & 0.003 & 17.326 & 0.009 &\nodata \\
2012 & 91 & 02:26:15.37 & +62:00:35.7 & 13.734 & 0.001 & 14.541 & 0.001 & 14.928 & 0.002 &\nodata \\
2013 & 60 & 02:26:09.59 & +62:00:58.3 & 14.425 & 0.001 & 15.404 & 0.001 & 15.689 & 0.003 &\nodata \\
2014 & 61 & 02:26:07.00 & +62:00:46.8 & 15.115 & 0.002 & 16.336 & 0.003 & 17.288 & 0.011 &\nodata \\
2016 & 59 & 02:26:01.91 & +62:01:12.2 & 15.438 & 0.001 & 16.451 & 0.002 & 16.739 & 0.005 &\nodata \\
2017 &\nodata & 02:25:52.29 & +62:00:09.9 & 17.441 & 0.005 & 18.837 & 0.011 & 19.427 & 0.047 &\nodata \\
2019 &\nodata & 02:25:46.88 & +62:01:09.2 & 16.185 & 0.002 & 17.437 & 0.004 & 18.118 & 0.016 &\nodata \\
2020 &\nodata & 02:25:49.17 & +62:01:17.5 & 17.661 & 0.008 & 19.176 & 0.017 & 19.909 & 0.083 &\nodata \\
2021 &\nodata & 02:25:52.16 & +62:01:28.9 & 15.939 & 0.002 & 17.360 & 0.003 & 18.015 & 0.015 &\nodata \\
2022 & 58 & 02:25:48.01 & +62:02:37.5 & 15.731 & 0.002 & 17.021 & 0.004 & 18.001 & 0.021 &\nodata \\
2023 & 57 & 02:26:01.55 & +62:02:54.5 & 14.518 & 0.001 & 15.391 & 0.002 & 15.561 & 0.004 &\nodata \\
2024 &\nodata & 02:26:03.49 & +62:02:32.5 & 16.976 & 0.007 & 18.315 & 0.014 & 19.146 & 0.067 &\nodata \\
2025 & 56 & 02:26:07.28 & +62:02:55.2 & 15.643 & 0.003 & 16.805 & 0.005 & 17.469 & 0.016 &\nodata \\
2026 &\nodata & 02:26:03.63 & +62:03:26.1 & 17.513 & 0.011 & 18.928 & 0.023 & 19.371 & 0.082 &\nodata \\
2027 &\nodata & 02:25:55.83 & +62:03:29.3 & 17.673 & 0.008 & 19.083 & 0.017 & 19.873 & 0.097 &\nodata \\
2028 &\nodata & 02:25:44.84 & +62:03:41.8 & 15.834 & 0.003 & 18.097 & 0.009 & 18.754 & 0.039 &\nodata \\
2029 &\nodata & 02:25:46.06 & +62:03:45.0 & 16.802 & 0.005 & 17.981 & 0.007 & 18.481 & 0.030 &\nodata \\
2030 &\nodata & 02:25:46.67 & +62:03:57.5 & 16.670 & 0.004 & 18.095 & 0.008 & 18.837 & 0.037 &\nodata \\
2032 &\nodata & 02:25:56.82 & +62:04:27.5 & 16.624 & 0.004 & 17.783 & 0.006 & 18.383 & 0.026 &\nodata \\
2033 &\nodata & 02:25:47.13 & +62:05:21.1 & 17.303 & 0.005 & 18.792 & 0.013 & 19.698 & 0.091 &\nodata \\
2036 & 55 & 02:26:08.24 & +62:04:53.4 & 14.445 & 0.001 & 15.436 & 0.002 & 15.734 & 0.005 &\nodata \\
2037 & 81 & 02:26:16.94 & +62:04:18.6 & 15.608 & 0.003 & 16.655 & 0.004 & 16.990 & 0.013 &\nodata \\
2038 &\nodata & 02:26:15.59 & +62:02:41.4 & 17.766 & 0.014 & 19.218 & 0.033 & 20.049 & 0.157 &\nodata \\
2039 & 54 & 02:26:10.11 & +62:06:12.3 & 16.027 & 0.002 & 16.983 & 0.004 & 17.185 & 0.010 &\nodata \\
2040 & 76 & 02:26:16.75 & +62:06:39.9 & 14.740 & 0.001 & 15.805 & 0.002 & 16.531 & 0.006 &\nodata \\
2041 & 77 & 02:26:18.83 & +62:06:27.7 & 13.191 & 0.001 & 14.070 & 0.001 & 14.379 & 0.001 &\nodata \\
2042 & 75 & 02:26:21.09 & +62:06:44.3 & 14.045 & 0.001 & 14.988 & 0.001 & 15.507 & 0.002 &\nodata \\
2043 &\nodata & 02:26:19.58 & +62:05:40.0 & 16.598 & 0.005 & 17.969 & 0.010 & 18.564 & 0.049 &\nodata \\
2044 &\nodata & 02:26:24.79 & +62:05:46.8 & 16.049 & 0.002 & 17.366 & 0.005 & 18.019 & 0.019 &\nodata \\
2045 &\nodata & 02:26:26.14 & +62:04:47.1 & 16.527 & 0.005 & 17.744 & 0.009 & 18.416 & 0.038 &\nodata \\
2046 &\nodata & 02:26:32.12 & +62:06:27.4 & 14.074 & 0.001 & 14.989 & 0.001 & 15.534 & 0.002 &\nodata \\
2048 & 80 & 02:26:38.29 & +62:06:11.1 & 13.326 & 0.001 & 14.419 & 0.001 & 14.889 & 0.002 &\nodata \\
2049 &\nodata & 02:26:33.69 & +62:05:34.9 & 16.589 & 0.003 & 17.838 & 0.007 & 18.465 & 0.032 &\nodata \\
2050 & 82 & 02:26:38.01 & +62:03:27.9 & 12.877 & 0.001 & 13.713 & 0.002 & 13.956 & 0.001 &\nodata \\
2051 &110 & 02:26:39.03 & +62:02:55.9 & 15.219 & 0.001 & 16.395 & 0.003 & 17.035 & 0.009 &\nodata \\
2052 &\nodata & 02:26:36.70 & +62:02:47.0 & 16.283 & 0.003 & 17.962 & 0.007 & 18.691 & 0.032 &\nodata \\
2053 & 83 & 02:26:34.32 & +62:01:52.8 & 14.372 & 0.001 & 15.285 & 0.001 & 15.361 & 0.003 &\nodata \\
2054 & 84 & 02:26:29.22 & +62:01:48.1 & 15.491 & 0.002 & 16.751 & 0.003 & 17.552 & 0.013 &\nodata \\
2055 &\nodata & 02:26:26.89 & +62:02:48.4 & 17.646 & 0.017 & 19.006 & 0.034 & 20.019 & 0.170 &\nodata \\
2056 &\nodata & 02:26:40.58 & +62:02:10.4 & 17.309 & 0.009 & 19.040 & 0.024 & 19.982 & 0.125 &\nodata \\
\enddata
\end{deluxetable}

\begin{deluxetable}{rrccccccccl}
\scriptsize
\tablewidth{0pt}
\tablecolumns{11}
\tablenum{\ref{t_phot}, {\it continued}}
\tablecaption{}
\startdata
3007 &\nodata & 02:26:44.60 & +62:00:48.0 & 16.942 & 0.005 & 18.424 & 0.011 & 19.095 & 0.042 &\nodata \\
3009 &\nodata & 02:26:47.81 & +62:01:06.9 & 17.568 & 0.006 & 19.322 & 0.019 & 20.619 & 0.167 &\nodata \\
3010 &\nodata & 02:26:50.89 & +62:00:45.5 & 17.136 & 0.005 & 18.778 & 0.013 & 19.651 & 0.066 &\nodata \\
3030 &\nodata & 02:26:52.69 & +62:04:11.9 & 17.669 & 0.012 & 19.132 & 0.026 & 20.092 & 0.141 &\nodata \\
3041 &109 & 02:26:45.29 & +62:03:07.6 & 13.114 & 0.002 & 14.846 & 0.001 & 15.145 & 0.003 & O9.7 Ia \\
3043 &\nodata & 02:26:50.85 & +62:02:53.4 & 17.559 & 0.011 & 19.253 & 0.029 & 19.657 & 0.100 &\nodata \\
3044 &\nodata & 02:26:51.92 & +62:02:28.0 & 17.680 & 0.011 & 19.035 & 0.023 & 19.698 & 0.100 &\nodata \\
4010 &\nodata & 02:26:45.69 & +61:59:50.8 & 16.632 & 0.004 & 18.170 & 0.010 & 18.710 & 0.040 &\nodata \\
4011 &\nodata & 02:26:46.41 & +61:59:47.3 & 16.631 & 0.004 & 18.455 & 0.013 & 19.330 & 0.068 &\nodata \\
4012 &111 & 02:26:46.66 & +62:00:26.1 & 13.932 & 0.002 & 14.908 & 0.001 & 14.993 & 0.003 & B2 V   \\
4060 &\nodata & 02:26:41.75 & +61:58:50.6 & 17.677 & 0.008 & 19.765 & 0.030 & 20.418 & 0.139 &\nodata \\
5009 &\nodata & 02:26:37.65 & +61:59:19.2 & 18.613 & 0.030 & 20.592 & 0.094 & 21.002 & 0.362 &\nodata \\
5014 &\nodata & 02:26:34.86 & +61:57:56.0 & 18.397 & 0.019 & 20.946 & 0.110 & 22.188 & 0.966 &\nodata \\
5037 &\nodata & 02:26:23.36 & +61:58:11.7 & 18.708 & 0.025 & 20.598 & 0.083 & 21.881 & 0.762 &\nodata \\
5041 &\nodata & 02:26:20.48 & +61:57:17.8 & 18.520 & 0.021 & 20.145 & 0.056 & 20.943 & 0.302 &\nodata \\
5069 &\nodata & 02:25:59.03 & +61:59:47.7 & 17.903 & 0.011 & 19.428 & 0.024 & 19.893 & 0.095 &\nodata \\
6020 &\nodata & 02:26:22.87 & +62:06:25.9 & 17.840 & 0.012 & 19.204 & 0.025 & 19.687 & 0.099 &\nodata \\
6024 &\nodata & 02:26:16.28 & +62:06:22.3 & 17.606 & 0.012 & 18.826 & 0.022 & 19.640 & 0.133 &\nodata \\
9039 &\nodata & 02:26:26.38 & +62:06:17.3 & 18.350 & 0.018 & 19.594 & 0.039 & 19.897 & 0.151 &\nodata \\
10003\tablenotemark{b} & 40 & 02:25:36.38 & +62:02:04.1 & 13.100 & 0.002 & 14.222 & 0.001 & 14.510 & 0.002 & G3 V   \\
10004 &\nodata & 02:25:34.59 & +62:01:40.1 & 16.471 & 0.011 & 17.751 & 0.018 & 18.035 & 0.045 &\nodata \\
10005 &\nodata & 02:25:36.70 & +62:01:18.9 & 16.153 & 0.007 & 17.421 & 0.012 & 18.154 & 0.042 &\nodata \\
10006 &\nodata & 02:25:37.75 & +62:01:19.1 & 16.853 & 0.016 & 18.143 & 0.024 & 18.447 & 0.060 &\nodata \\
10020 &33 & 02:25:40.67 & +62:04:19.8 & 15.102 & 0.002 & 16.029 & 0.003 & 16.171 & 0.006 &\nodata \\
10021 &35 & 02:25:34.79 & +62:03:28.9 & 16.399 & 0.006 & 17.485 & 0.009 & 17.708 & 0.025 &\nodata \\
\enddata
\tablenotetext{a}{Photometry suspect, yielding unphysical parameters; star omitted from analysis.}
\tablenotetext{b}{Foreground star.}
\tablenotetext{c}{Spectral type from Mathys (1989).}
\end{deluxetable}

\section{Discussion and Conclusion}

We now return to the question of hierarchical triggered star formation
in the entire W3/W4 complex.

The age of the 230-pc Perseus chimney/superbubble feature is
somewhat problematic.  Dennison et al. (1997) roughly estimate it to be
6 -- 10 Myr old, while Basu et al. (1999) find a much younger
age of $\sim$2.5 Myr, from a more detailed semi-analytic investigation.
Basu et al. solve simultaneously for the age and scale height of the
ambient gas, and it is important to note that this young age estimate
is largely due to the surprisingly small, 25-pc, Galactic scale height
of the ambient gas that they find.  Reynolds et al. (2001) argue that
the largest, 1300-pc loop associated with the region requires at least
10 -- 20 Myr to form.  

The stellar population at the center of W4, at
the apex of the chimney/superbubble system, is extremely young.  The
central OB association, IC 1805, is around 1 -- 3 Myr old, as
determined by Massey et al. (1995a) from a detailed study including
spectral classifications of the earliest-type stars.  This age is
consistent with that determined for the wider population, Cas~OB6 /
OCl~352 (e.g., Normandeau et al. 1996; Mathys 1987; Llorente de
Andr\`es et al. 1982).

Dennison et al. (1997) suggest that an earlier generation of star
formation, preceding the birth of IC 1805, was responsible for
forming the 230-pc Perseus superbubble.  Indeed, Carpenter et
al. (2000) independently suggest the existence of an older
star-forming episode to explain the large number and distribution of
molecular cloudlets and globules scattered in the wider W4 area.
They speculate that these
molecular globules are the remnants of a giant molecular cloud that
was dispersed by the photodissociation, stellar winds, and supernovae
of the resulting star-forming region.  

IC 1805, at the center of W4, represents a more recent
generation.  The young, 1 -- 3 Myr age found by Basu et al. (1999) for
the Dennison superbubble is consistent with the age of IC 1805,
suggesting that this association is responsible for that shell.  
The unusually small scale height that is linked to this young age
by Basu et al. is also beautifully consistent with the existence of an
earlier generation and earlier superbubble in the same region, which
would have cleared out much of the ambient ISM, in particular across
Galactic latitude.  This naturally points to the larger,
1300-pc shell reported by Reynolds et al. (2001), that they estimate to
be 10 -- 20 
Myr old.  Note that if a normal ISM scale height is assumed, the age
of the Dennison superbubble would be older, around 6 -- 10 Myr old, as
estimated by those authors, regardless of the age of IC 1805.  Thus,
it is a fairly robust result that the superbubble activity associated
with W4 is at least 6 -- 20 Myr old.

From Figure~\ref{f_hrd}, we derived an age for IC~1795, in the center
of the W3 superbubble, of 3 -- 5 Myr.  This age is consistent with the
location of IC~1795 in the central, cleared area of the molecular
shell (Figure~\ref{f_w3}), while younger, on-going star formation, is
taking place in the ultracompact \hii\ regions on the edge of that
shell, in W3-North, W3-Main, and W3-OH.   These youngest star-forming
regions show individual sources embedded in their molecular gas cores,
which are of order $10^4$ to $10^5$ years old (e.g. Megeath et
al. 1996; Tieftrunk et al. 1998).   
Therefore, the entire W3/W4 complex does indeed show an age sequence
consistent with hierarchical triggering of star formation:  The
first generation corresponds to the 1300~pc loop, and possibly the
Perseus superbubble, along with the progenitor stellar population;
these triggered the formation of IC~1795 on the western edge; which in
turn triggered the formation of W3-North, W3-Main, and W3-OH at the
edges of its shell (Figures~\ref{f_dennison}--\ref{f_w3}).  IC 1805,
at the apex of the W4 shells by projection, is 
a young, 1 -- 3 Myr old association that appears not to be responsible
for triggering the creation of IC~1795.  It is possible that the
formation of IC 1805 was itself triggered by the earliest generation,
and possibly it, too, is located on the edge of the shell, although
projected toward the center in our line of sight.

In contrast to two-generation superbubble systems, this 
three-generation configuration is extremely difficult to attribute
to coincidental sequential star formation.  Thus, the W3/W4 complex
provides some of the strongest evidence to date that superbubble
activity and mechanical feedback from massive stars is indeed a
mechanism for triggering star formation.  Additional examples of
confirmed multi-generation star formation can further confirm this.

\acknowledgments

We thank Phil Massey for discussions on the photometry, and
Brian Skiff for cross-referencing the stars in the literature.  
MSO and GLW acknowledge support from the National Science Foundation, 
grant AST-0204853.  KK participated through the Research Experience
for Undergraduates program at Northern Arizona University, also
supported by the National Science Foundation.  Much of this work was
done by MSO at the Space Telescope Science Institute and Lowell
Observatory.  




\appendix

\medskip
Here we present the complete, new stellar dataset for the greater W3 region.
Table~\ref{t_photall} presents our $UBV$ photometry for stars observed in
the region covered by Figure~\ref{f_mosaic}.  (The complete version of 
Table~\ref{t_photall} is available in the on-line edition).
The first two columns give ID numbers assigned by this work and Ogura
\& Ishida (1976), respectively, columns 3 and 4 give the celestial
coordinates for epoch J2000.0, and columns 5 -- 10 list the $V$, $B$, and
$U$ magnitudes with corresponding formal (non-systematic)
uncertainties.  Finally, column 11 gives the number of observations
that were averaged to obtain the final values, and column 12 gives the
spectral classification resulting from our spectroscopic data.

\medskip
{\noindent\it Special notes:  }

Star 2001 (OI 89 = BD +61 411):  Formal uncertainties $ < 0.001$ in all
three bands.  This star was classified as \break O6.5 V((f)) by Mathys (1989).

Star 9018 (OI 71):  $V$ measurement saturated; listed value is adopted
from Ogura \& Ishida (1976).

Star 2046:  Resolved into two stars, OI 78 and OI 79, by Ogura \&
Ishida (1976).

\begin{deluxetable}{rrcccccccccl}
\footnotesize
\tablewidth{0pt}
\tablecolumns{12}
\tablecaption{Photometry and Spectral Types for the W3 region \label{t_photall}}
\tablehead{
\colhead{ID} & {OI} & {RA(J2000.0)} & {Dec(J2000.0)} &
\colhead{$V$} & {$V$ err} & \colhead{$B$} & {$B$ err} & 
\colhead{$U$} & {$U$ err} & n-obs & Sp Type }
\startdata
1007 & 112 & 2:26:41.31 & 61:59:24.2 & 14.318 & 0.001 & 15.256 & 0.001 & 15.299 & 0.002  & 3 & B1-2 V \\
1008 & \nodata & 2:26:34.63 & 61:59:10.6 & 15.312 & 0.001 & 16.947 & 0.003 & 17.753 & 0.011  & 3 & \nodata \\
1009 & \nodata & 2:26:38.45 & 61:59:26.2 & 18.138 & 0.013 & 19.585 & 0.028 & 20.080 & 0.120  & 2 & \nodata \\
1010 & \nodata & 2:26:40.83 & 61:58:11.7 & 17.233 & 0.005 & 18.736 & 0.011 & 18.000 & 0.023  & 3 & \nodata \\
1011 & \nodata & 2:26:34.54 & 61:58:16.6 & 18.124 & 0.013 & 20.069 & 0.042 & 20.883 & 0.235  & 2 & \nodata \\
1012 & \nodata & 2:26:34.69 & 61:57:32.9 & 17.853 & 0.010 & 19.403 & 0.023 & 20.086 & 0.121  & 2 & \nodata \\
1013 & \nodata & 2:26:35.17 & 61:57:11.0 & 16.616 & 0.003 & 17.989 & 0.006 & 18.891 & 0.030  & 3 & \nodata \\
1014 & \nodata & 2:26:39.27 & 61:57:06.2 & 17.501 & 0.006 & 19.211 & 0.015 & 20.046 & 0.104  & 3 & \nodata \\
1015 & 93  & 2:26:33.94 & 61:56:57.2 & 14.739 & 0.001 & 15.757 & 0.001 & 16.043 & 0.003  & 3 & \nodata \\
1016 & \nodata & 2:26:33.30 & 61:56:58.6 & 15.584 & 0.001 & 16.914 & 0.003 & 16.923 & 0.008  & 3 & \nodata \\
\enddata
\tablecomments{Table~\ref{t_photall} is presented in its entirety in the electronic edition
of the Astronomical Journal.  A portion is shown here for guidance regarding its form and content.}
\end{deluxetable}
 
\clearpage
\vfill\eject

\end{document}